\documentclass[hidelinks,journal]{ieeeconf}      

\IEEEoverridecommandlockouts                              

\usepackage{xr}

\newcommand{\omitted}[1]{}

\title{\fontsize{22}{22} \selectfont Safe Control of Partially-Observed Linear Time-Varying \\ Systems with Minimal Worst-Case Dynamic Regret}

\author{Hongyu Zhou and Vasileios Tzoumas$^\star$\vspace{-6mm}
	\thanks{$^\star$Department of Aerospace Engineering, University of Michigan, Ann Arbor, MI 48109 USA;  {\tt\footnotesize \{zhouhy, vtzoumas\}@umich.edu}}
}

\usepackage{cite}

\usepackage{comment}
\usepackage{siunitx}
\usepackage{relsize}
\usepackage{ifthen}
\usepackage[colorinlistoftodos]{todonotes}

\usepackage[caption=false]{subfig}

\usepackage[vlined,ruled,linesnumbered]{algorithm2e}
\usepackage{graphics} 
\usepackage{rotating}
\usepackage{color}
\usepackage{enumerate}
\usepackage[T1]{fontenc}
\usepackage{psfrag}
\usepackage{epsfig} 
\usepackage{booktabs}
\usepackage{graphicx,url}
\usepackage{multirow}
\usepackage{array}
\usepackage{latexsym}
\usepackage{amsfonts}
\usepackage{amsmath}

\usepackage{amssymb}
\usepackage{xstring}
\usepackage{algorithmic}
\usepackage{multirow}
\usepackage{xcolor}
\usepackage{prettyref}
\usepackage{flexisym}
\usepackage{bigdelim}
\usepackage{breqn} 
\usepackage{listings}
\let\labelindent\relax
\usepackage{xspace}
\usepackage{bm}
\graphicspath{{./figures/}}
\usepackage{tikz}
\usetikzlibrary{matrix,calc}
\usepackage{lipsum}
\usepackage{mdwlist}

\let\itemize\undefined
\makecompactlist{itemize}{stditemize}
\usepackage{enumitem}
\usepackage{caption}
\usepackage{epstopdf}
\usepackage{amsthm}
\usepackage{mathtools}

\makeatletter
\let\NAT@parse\undefined
\makeatother
\usepackage{hyperref}
\usepackage{cleveref}

\hypersetup{%
  colorlinks=true,
  linkcolor=blue,
  filecolor=magenta,      
  urlcolor=black,
  citecolor=red,
  linkbordercolor={0 0 1}
}

\usepackage{float}

\newtheorem{theorem}{Theorem}
\newtheorem{problem}{Problem}

\newtheorem{assumption}{Assumption}
\newtheorem{definition}{Definition}
\newtheorem{proposition}{Proposition}

\newcommand{\bdmath}{\begin{dmath}}
\newcommand{\edmath}{\end{dmath}}
\newcommand{\beq}{\begin{equation}}
\newcommand{\eeq}{\end{equation}}
\newcommand{\bdm}{\begin{displaymath}}
\newcommand{\edm}{\end{displaymath}}
\newcommand{\bea}{\begin{eqnarray}}
\newcommand{\eea}{\end{eqnarray}}
\newcommand{\beal}{\beq \begin{array}{lll}}
\newcommand{\eeal}{\end{array} \eeq}
\newcommand{\beas}{\begin{eqnarray*}}
\newcommand{\eeas}{\end{eqnarray*}}
\newcommand{\ba}{\begin{array}}
\newcommand{\ea}{\end{array}}
\newcommand{\bit}{\begin{itemize}}
\newcommand{\eit}{\end{itemize}}
\newcommand{\ben}{\begin{enumerate}}
\newcommand{\een}{\end{enumerate}}

\newcommand{\calD}{{\cal D}}

\newcommand{\calH}{{\cal H}}

\newcommand{\calK}{{\cal K}}

\newcommand{\calQ}{{\cal Q}}
\newcommand{\calR}{{\cal R}}

\definecolor{myblue}{RGB}{65 105 225}

\newcommand{\hide}[1]{}

\newcommand{\hiddenText}{{\color{gray} hidden text.}}
\newcommand{\hideWithText}[1]{\hiddenText}

\newcommand{\myK}{\mathbf{\calK}}
\newcommand{\myPhi}{\mathbf{\Phi}}

\newcommand{\ie}{\emph{i.e.},\xspace}
\newcommand{\eg}{\emph{e.g.},\xspace}

\newcommand{\myParagraph}[1]{{\bf #1.}\xspace}

\PassOptionsToPackage{end}{algorithmic}

\renewcommand{\baselinestretch}{0.98}

\begin{document}

\maketitle

\thispagestyle{empty}
\pagestyle{empty}

\begin{abstract}
We present safe control of partially-observed linear time-varying systems in the presence of unknown and unpredictable process and measurement noise. We introduce a control algorithm that minimizes \textit{dynamic regret}, \ie that minimizes the suboptimality against an optimal clairvoyant controller that knows the unpredictable future a priori. Specifically, our algorithm minimizes the worst-case {dynamic regret} among all possible noise realizations given a worst-case total noise magnitude.  To this end, the control algorithm accounts for three key challenges: {safety constraints}; {partially-observed time-varying systems}; and {unpredictable process and measurement noise}. We are motivated by the future of autonomy where robots will autonomously perform complex tasks despite unknown and unpredictable disturbances leveraging their on-board control and sensing capabilities. To synthesize our minimal-regret controller, we formulate a constrained semi-definite program based on a System Level Synthesis approach for partially-observed time-varying systems. We validate our algorithm in simulated scenarios, including trajectory tracking scenarios of a hovering quadrotor collecting GPS and IMU measurements.
Our algorithm is observed to have better performance than either or both the $\mathcal{H}_2$ and $\mathcal{H}_\infty$ controllers, demonstrating a \textit{Best of Both Worlds} performance.
\end{abstract}

\vspace{-3mm}
\section{Introduction}\label{sec:Intro}

In the future, robots will be leveraging their on-board control and sensing capabilities to complete tasks such as {package delivery}~\cite{ackerman2013amazon}, {transportation}~\cite{rajendran2020air}, and {disaster response}~\cite{rivera2016post}.
To complete such complex tasks, the robots need to \textit{reliably} overcome a series of key challenges:
\begin{itemize}[leftmargin=9pt]
    \item \textbf{Challenge I: Safety Constraints.} Robots need to ensure their own safety and the safety of their surroundings.  For example,  robots often need to ensure that they follow prescribed collision-free trajectories or that their control effort is kept under prescribed levels. Such safety requirements take the form of state and control input constraints and make {planning control inputs computationally hard~\cite{rawlings2017model,borrelli2017predictive}.}
    
    \item \textbf{Challenge II: Partially-Observed Time-Varying System.} 
    Robots often lack full-state information feedback for control.  Instead, they base their control on sensing capabilities that are governed and time-varying measurement models.  For example, self-driving robots in indoor environments often base their control on range sensors which provide relative-distance measurements to known landmarks~\cite{thrun2002probabilistic}.  
    The measurement model is time-varying since range measurements depend on the relative position of the robot to the landmarks. Accounting for such models typically requires linearization~\cite{brown1997introduction}, which adds to the hardness of computing safe control inputs.
    
    \item \textbf{Challenge III: Unknown and Unpredictable Process and Measurement Noise.} The robots' dynamics and measurements are often disturbed by unknown and unstructured noise, which is not necessarily Gaussian.  For example, aerial and marine vehicles often face unpredictable winds and waves~\cite{faltinsen1993sea, sapsis2021statistics}.  But the current control algorithms primarily rely on known or Gaussian-structured noise, compromising thus the robots' ability to ensure safety
    against unknown and unpredictable noise~\cite{aastrom2012introduction,berkenkamp2019safe}.
\end{itemize}

The above challenges motivate the development of safe control algorithms for partially-observable linear time-varying systems, guaranteeing near-optimal control performance against unknown and unpredictable noise.  

\myParagraph{Related Work} The current control algorithms either consider (i) no safety constraints, or (ii) fully-observed systems or partially-observed systems but with time-\underline{in}variant measurement models, or (iii) known stochastic models about the process and measurement noise, \eg Gaussian noise.

We next review the literature by first reviewing \textit{online learning for control} algorithms, \ie algorithms that select control inputs based on past information only~\cite{hazan2020nonstochastic,agarwal2019online,li2021online,simchowitz2020improper,gradu2020adaptive}, and then by reviewing \textit{robust control} algorithms that select inputs based on simulating the future system dynamics across a {lookahead} horizon~\cite{goel2020regret,sabag2021regret,goel2021regret,martin2022safe,didier2022system}:

\paragraph{Online Learning for Control} The algorithms performing online learning for control make no assumptions about the noise's stochasticity~\cite{shalev2012online,hazan2016introduction}, aiming to address the Challenge III.  They assume the noise can evolve arbitrarily, subject to a given upper bound on its magnitude.  The upper bound ensures problem feasibility, and tunes the algorithms' response to the nevertheless unknown noise.   

The online learning algorithms prescribe control policies by optimizing feedback control gains based on past information only,  guaranteeing performance by bounding static or  dynamic regret: \textit{static regret}~\cite{shalev2012online, hazan2016introduction} captures the algorithms' suboptimality against optimal {static} policies that know the future a priori, \ie control policies with \textit{time-\underline{in}variant} control gains~\cite{hazan2020nonstochastic,agarwal2019online,li2021online,simchowitz2020improper}; and \textit{dynamic regret}~\cite{zinkevich2003online} captures the algorithm's suboptimality against optimal {dynamic} policies that know the future a priori, \ie control policies with \textit{time-varying} control gains~\cite{gradu2020adaptive}.  

The current static-regret algorithms consider no safety constraints, with the exception of \cite{li2021online},
and apply to fully-observed systems or partially-observed systems with time-\underline{in}variant measurement models only; and the current dynamic-regret algorithms also ignore safety constraints, and apply to fully-observed systems only. All in all, although the current online learning algorithms address unpredictable noise, they assume no safety constraints and cannot apply to partially-observed time-varying systems. 

\paragraph{Robust Control}
The classical $\mathcal{H}_2$ and $\mathcal{H}_\infty$ control algorithms~\cite{hassibi1999indefinite} assume Gaussian noise and bounded worst-case noise, respectively. Particularly, $\calH_2$ is optimal under Gaussian noise but it can thus underperform under non-Gaussian noise. $\calH_\infty$ on the other hand can be conservative.

For this reason, recent robust control algorithms aim to address the challenge of unpredictable noise by focusing on convex optimization techniques that guarantee \textit{regret optimality}, \ie minimal {worst-case dynamic-regret} among all noise realizations subject to a given total noise magnitude.~\cite{goel2020regret,sabag2021regret} focus on fully-observed systems and~\cite{goel2021regret} on partially-observed systems.  But these algorithms consider no safety constraints. Instead, \cite{didier2022system,martin2022safe} provide a regret-optimal control algorithm that accounts for safety constraints.  However, they focus on fully-observed systems only. 


\myParagraph{Contributions} 
In this paper, we provide an algorithm with dynamic regret guarantees for the safe control of partially-observed linear time-varying systems. The algorithm prescribes an output-feedback control input, guaranteeing minimum worst-case dynamic regret among all noise realizations. It is a robust control algorithm, selecting inputs based on simulating the dynamics across a lookahead horizon.
To this end, we formalize the problem of \textit{Safe Control of Partially-Observed Linear Time-Varying Systems with Minimal Worst-Case Regret} (\Cref{prob_safe_obs_reg_opt}).
%

Our analysis builds on~\cite{wang2019system, anderson2019system,martin2022safe}:
    (i) We prove that the output-feedback control gains are the solution of a constrained Semi-Definite Program (SDP); our SDP approach handles partially-observed systems, generalizing from the fully-observed case presented in~\cite{martin2022safe}. (ii) We use a \textit{System Level Synthesis} (SLS) method that extends the current SLS methods~\cite{wang2019system, anderson2019system} to partially-observable time-varying systems, presenting necessary and sufficient conditions for the existence of a causal safe output-feedback control policy.

\myParagraph{Numerical Evaluations} We validate our algorithm in simulated scenarios of partially-observed linear time-varying systems, including trajectory tracking scenarios of a hovering quadrotor collecting GPS and Inertial Measurement Unit (IMU) measurements (\Cref{sec:experiments}). 
We compare our algorithm with the safe $\mathcal{H}_2$ and $\mathcal{H}_\infty$ control algorithms \cite{martin2022safe} under diverse process and measurement noises.

Our algorithm demonstrates a \textit{Best of Both Worlds} performance across all simulations and test types of noise, performing on average better than at least one of the $\calH_2$ and $\calH_\infty$ controllers. That is, our algorithm demonstrates robustness across all tested types of noise, being the best or the second best among $\mathcal{H}_2$ and $\mathcal{H}_\infty$, an advantageous performance capacity when the type of noise is unknown a priori and unpredictable. Instead, $\calH_2$ is the worst against worst-case noise and $\calH_\infty$ is the worst against Gaussian. 

\myParagraph{Organization}
\Cref{sec:problem} formulates the problem of \textit{safe control of partially-observed linear time-varying systems with minimal worst-case dynamic regret guarantees}. \Cref{sec:method} develops the control algorithm. \Cref{sec:experiments} presents the numerical evaluation. \Cref{sec:con} concludes the paper.  The Appendix contains all proofs.

\section{Problem Formulation}\label{sec:problem}

We formulate the problem of \textit{Safe Control of Partially-Observed Linear Time-Varying Systems with Minimal Worst-Case Regret} (\Cref{prob_safe_obs_reg_opt}). We use the framework:

\myParagraph{Partially-Observed System}
We consider partially-obser- ved Linear Time-Varying (LTV) systems of the form
\begin{equation}
    \begin{aligned}
        x_{t+1} &= A_{t} x_{t} + B_{t} u_{t} + w_{t}, \quad t \in \{0, \ldots, T-1\}, \\
        y_{t} &= C_{t} x_{t} + e_{t},
    \end{aligned}
    \label{eq_LTV}
\end{equation}
where $t$ is the time index, $T$ is a time horizon of interest, $x_{t} \in \mathbb{R}^{d_x}$ is the system's state, $u_{t} \in \mathbb{R}^{d_u}$ is the control input, $w_{t} \in \mathbb{R}^{d_x}$ is the process noise, $y_{t} \in \mathbb{R}^{d_y}$ is the measurement, and $e_{t} \in \mathbb{R}^{d_y}$ is the measurement noise. 

We henceforth denote:
\begin{itemize}[leftmargin=9pt]
    \item $\mathbf{x} \,\triangleq\,\left[x_{0}^\top, x_{1}^\top, \ldots, x_{T-1}^\top\right]^\top$, \ie $\mathbf{x}$ is the state trajectory across the time horizon $T$;
    \item $\mathbf{u}$, $\mathbf{y}$, and $\mathbf{e}$ are defined correspondingly to $\mathbf{x}$;
    \item $\mathbf{w} \,\triangleq\,\left[x_{0}^\top, w_{0}^\top, \ldots, w_{T-2}^\top\right]^\top$, \ie is the process noise trajectory till time $T-2$ appended with the initial condition.
\end{itemize} 

\begin{assumption}[Known System] \label{assumption_sys}
The initial condition $x_0$ and the matrices $A_{t}$, $B_{t}$, and $C_{t}$ for all $t$ are known.
\end{assumption} 

\begin{assumption}[Bounded Noise] \label{assumption_noise}
The process and measurement noise $\mathbf{w}_t$ and $\mathbf{e}_t$ are constrained in known compact polytopes that contain a neighborhood of the origin: \ie we are given $\mathbb{W}~\triangleq~\left\{\mathbf{w} \in \mathbb{R}^{d_x T}: \mathbf{H}_{w} \mathbf{w} \leq \mathbf{h}_{w}\right\}$ and $\mathbb{E}~\triangleq\left\{\mathbf{e} \in\right.$ $\left.\mathbb{R}^{d_y T}: \mathbf{H}_{e} \mathbf{e} \leq \mathbf{h}_{e}\right\}$ for given matrices $\mathbf{H}_{w}$, $\mathbf{H}_{e}$, $\mathbf{h_w}$ and $\mathbf{h_e}$.
\end{assumption}

Per \Cref{assumption_noise}, we assume no stochastic model for the noise.  Specifically, the noise may even be adversarial, subject to the bounds prescribed by $\mathbb{W}$ and $\mathbb{E}$. 

\myParagraph{Safety Constraints} We consider the states and control inputs must {satisfy polytopic constraints of the form}
\begin{equation}
    \mathbf{H}\left[\begin{array}{c}
    \mathbf{x} \\
    \mathbf{u}
    \end{array}\right] \leq \mathbf{h}, \ \forall \mathbf{w} \in \mathbb{W}, \ \forall\mathbf{e} \in \mathbb{E},
    \label{eq_safety_constraints}
\end{equation}
for given matrices $\mathbf{H}$ and $\mathbf{h}$.

\myParagraph{Output-Feedback Control Policy} We consider the following output-feedback control policy:
\begin{equation}
    u_{t} = \sum_{k=0}^{t} K_{t,k} y_{k},  \quad t \in \{0, \ldots, T-1\},
    \label{eq_ut}
\end{equation}
where $K_{t,k}$ are control gains to be designed in this paper.  

\myParagraph{Control Performance Metric} We design the output-feed- back control gains $K_{t,k}$ to ensure both safety and a control performance comparable to an optimal clairvoyant policy that selects control inputs knowing the future noise realizations $\mathbf{w}$ and $\mathbf{e}$ a priori.  In this paper, we consider that the clairvoyant policy minimizes a cost of the form
\begin{equation}\label{eq:cost-basic}
    \operatorname{cost}(\mathbf{w},\mathbf{e}, \mathbf{u})~\triangleq~\mathbf{x}^\top \mathcal{Q}\mathbf{x}+\mathbf{u}^\top\mathcal{R}\mathbf{u},
\end{equation}
where $\mathcal{Q} \succeq 0$ and $\mathcal{R} \succ 0$, and $\mathbf{x}$ is a function of the control input sequence $\mathbf{u}$ and the noise $\mathbf{w}$ and  $\mathbf{e}$ per \cref{eq_LTV}.  $\mathcal{Q}$ and $\mathcal{R}$ are assumed symmetric, without loss of generality. Then, the suboptimality of any (causal) control sequence $\mathbf{u}$ that is unaware of the noise realization $\mathbf{w}$ and $\mathbf{e}$ is captured by the
\begin{equation}    \label{eq_reg}
        \operatorname{{regret}}_T(\mathbf{w},\mathbf{e},\mathbf{u})\triangleq
        \operatorname{cost}(\mathbf{w},\mathbf{e},\mathbf{u})-\min _{\mathbf{u}^{\prime} \in \mathbb{R}^{d_u T}} \operatorname{cost}\left(\mathbf{w},\mathbf{e}, \mathbf{u}^{\prime}\right),
    \end{equation}
where $\min _{\mathbf{u}^{\prime} \in \mathbb{R}^{d_u T}} \operatorname{cost}\left(\mathbf{w},\mathbf{e}, \mathbf{u}^{\prime}\right)$ is the cost achieved by the optimal clairvoyant control policy. 



In this paper, we design the output-feedback control gains $K_{t,k}$  to minimize the \textit{worst-case dynamic regret} among all feasible noise realizations per \Cref{assumption_noise}.

\begin{definition}[Worst-Case Dynamic Regret~\cite{goel2021regret}] \label{def_dyn_regret}
Denote by $r$ the minimum radius of a ball in $\mathbb{R}^{d_x T + d_u T}$ that encircles the noise's domain sets $\mathbb{W}$ and $\mathbb{E}$. Then, 
\begin{equation}
        \operatorname{{worst-case-regret}}_T(\mathbf{u})\triangleq\max _{\|\mathbf{w}\|_2^2\,+\,\|\mathbf{e}\|_2^2\,\leq\, r^2
        } \   \operatorname{regret}_T(\mathbf{w},\mathbf{e}, \mathbf{u}).
    \label{eq_dyn_reg}
    \end{equation}
\end{definition}

That is, \cref{eq_dyn_reg} is the worst-case dynamic regret among all noise realizations with maximum feasible total magnitude.

\myParagraph{Problem Definition}  In this paper, we focus on: 

\begin{problem}[Safe Control of Partially-Observed Linear Time-Varying Systems with Minimal Worst-Case Regret] \label{prob_safe_obs_reg_opt}
Find control gains $K_{t,\;k}$ such that the output-feedback control policy in \cref{eq_ut} guarantees (i) the safety of the partially-observed LTV system in \cref{eq_LTV} and (ii) minimal worst-case dynamic regret. Formally,
\begin{subequations}
\begin{align}
     \underset{\scriptsize{\begin{array}{c}
         K_{t,\;k}, \\
         \forall (t,\,k)\in\{0,1,\ldots,T\}^2
    \end{array}}
    }{\operatorname{\emph{min}}} &\;\operatorname{worst-case-regret}_T(\mathbf{u})\\
   \text{\emph{subject to}}  \!\hspace{-1mm}\quad &\;\;\text{the system in~\cref{eq_LTV}};\\
    &\;\;\text{the safety constraints in~\cref{eq_safety_constraints}};\\
    &\;\;\text{the control policy in~\cref{eq_ut}}.
\end{align}
\label{eq_prob_safe_obs_reg_opt}
\end{subequations}
\end{problem}

\vspace{-10mm} 

\section{Algorithm for {\Cref{prob_safe_obs_reg_opt}}}\label{sec:method}

We present the algorithm for \Cref{prob_safe_obs_reg_opt} (\Cref{alg:safe_obs_reg_opt}). The algorithm solves \Cref{prob_safe_obs_reg_opt} via an equivalent SDP reformulation.  We present the SDP reformulation in \Cref{subsec:main-result} (\Cref{theorem_safe_obs_reg_opt}).  To prove the SDP reformulation, we first present an SLS approach for partially-observed linear time-varying systems in \Cref{subsec:sls} (\Cref{prop_SLS_LTVwithMeasurements}).

\vspace{-3mm}
\subsection{Preliminary: System Level Synthesis for Partially-Observable LTV Systems} \label{subsec:sls}
We use the SLS approach to partially-observed LTV systems. Particularly, given a desired state trajectory $\mathbf{x}$, we present necessary and sufficient conditions for the existence of a control policy $u_t$ per \cref{eq_ut}. Equivalently, we show necessary and sufficient conditions for the existence of control gains $K_{t,k}$ such that $K_{t,k}=0$ when $k>t$. The conditions take the form of linear constraints and thus enable the computation of the $K_{t,k}$ within an SDP reformulation of \Cref{prob_safe_obs_reg_opt} (\Cref{subsec:main-result}).

To the above ends, we use the notation:
\begin{itemize}[leftmargin=9pt]
    \item $\mathbf{\mathcal{Z}}$ denotes the block-matrix downshift operator, \ie 
    \begin{equation}
    \mathbf{\mathcal{Z}}\,\triangleq\,\left[\begin{array}{cccc}
                    \mathbf{0} & \mathbf{0} & \ldots & \mathbf{0}\\
                    \mathbf{I} & \ddots & \ddots & \vdots\\
                    \vdots & \ddots & \ddots & \mathbf{0}\\
                    \mathbf{0} & \ldots & \mathbf{I}  & \mathbf{0}
                    \end{array}\right];
    \end{equation}
    \item $\mathcal{A}$, $\mathcal{B}$, and $\mathcal{C}$ are the diagonal block-matrices whose block diagonal is the (partial) trajectory of the corresponding system, input, and measurement matrix, \ie
\begin{equation}
    \begin{aligned}
        \mathbf{\mathcal{A}}&\,\triangleq\,\operatorname { blkdiag }\left(A_{0}, A_{1}, \ldots, A_{T-2}, 0_{d_x \times d_x} \right); \nonumber\\
        \mathbf{\mathcal{B}}&\,\triangleq\,\operatorname{blkdiag}\left(B_{0}, B_{1}, \ldots, B_{T-2}, 0_{d_x \times d_u} \right); \nonumber \\
        \mathbf{\mathcal{C}}&\,\triangleq\,\operatorname { blkdiag }\left(C_{0}, C_{1}, \ldots, C_{T-1}\right);\nonumber
    \end{aligned}
    \label{eq_ABC_block}
\end{equation}
    \item $\mathcal{K}$ is the lower triangular block-matrix such that \cref{eq_ut} takes the form $\mathbf{u}=\calK\mathbf{y}$, \ie
    
    \begin{equation}\label{eq:calK}
        \mathbf{\mathcal{K}} \,\triangleq\, \left[\begin{array}{cccc}
K_{0,0} & 0_{d_u \times d_y} & \ldots & 0_{d_u \times d_y}\\
K_{1,0} & K_{1,1} & \ddots & \vdots\\
\vdots & \vdots & \ddots & 0_{d_u \times d_y}\\
K_{T-1, 0} & K_{T-1, 1} & \ldots  & K_{T-1, T-1}
\end{array}\right].
    \end{equation}
\end{itemize} 
Equations \eqref{eq_LTV} and \eqref{eq_ut} now take the form 

\hspace{-.35cm}\begin{minipage}{\columnwidth}
\begin{subequations}
\vspace*{-2mm}
    \begin{align}  
        \mathbf{x} &= \mathcal{Z} \mathcal{A} \mathbf{x}+ \mathcal{Z} \mathcal{B} \mathbf{u}+\mathbf{w};    \label{eq_system_compact_x} \\
        \mathbf{y} &= \mathcal{C} \mathbf{x} + \mathbf{e}; \label{eq_system_compact_y}\\
        \mathbf{u} &= \mathcal{K} \mathbf{y}, \label{eq_system_compact_u} 
    \end{align}\label{eq_system_compact}
    \vspace{-2mm}
\end{subequations}
\end{minipage}

\noindent which can be equivalently written as 
\begin{equation}
    \left[\begin{array}{c}
        \mathbf{x} \\
        \mathbf{u}
    \end{array}\right] = 
         \left[\begin{array}{cc}
            \mathbf{\Phi}_{xw} & \mathbf{\Phi}_{xe} \\
            \mathbf{\Phi}_{uw} & \mathbf{\Phi}_{ue}
             \end{array}\right] 
             \left[\begin{array}{c}
                \mathbf{w} \\
                \mathbf{e}
             \end{array}\right],     \label{eq_closedloop_mapping}
\end{equation}
\noindent where:
    \vspace{-2mm}
    \begin{subequations}
    \begin{align}   
        & \mathbf{\Phi}_{xw} = \left( \mathbf{I} - \mathcal{Z} \mathcal{A} - \mathcal{Z} \mathcal{B} \mathcal{K} \mathcal{C}\right)^{-1}, \label{eq_Phi_xw}\\
        & \mathbf{\Phi}_{xe} = \mathbf{\Phi}_{xw} \mathcal{Z} \mathcal{B} \mathcal{K}, \label{eq_Phi_xe}\\
        & \mathbf{\Phi}_{uw} = \mathcal{K} \mathcal{C} \mathbf{\Phi}_{xw}, \label{eq_Phi_uw}\\
        & \mathbf{\Phi}_{ue} = \mathcal{K} \mathcal{C} \mathbf{\Phi}_{xe} + \mathcal{K} \label{eq_Phi_ue}.
    \end{align} \label{eq_Phi}
    \vspace{-5mm}
\end{subequations}

\noindent We refer to the matrix $$\mathbf{\Phi}\triangleq \left[\begin{array}{cc}
                       \mathbf{\Phi}_{xw} & \mathbf{\Phi}_{xe} \\
                       \mathbf{\Phi}_{uw} & \mathbf{\Phi}_{ue}
                       \end{array}\right]$$
in \cref{eq_closedloop_mapping} as the \textit{response matrix}.

Given the lower triangular block-matrix $\myK$, \cref{eq_closedloop_mapping} captures how the noise trajectories $(\mathbf{w}, \mathbf{e})$ result to the control input trajectory $\mathbf{u}$ and, all in all, to the state trajectory $\mathbf{x}$.  Particularly, \mbox{$\mathbf{\Phi}_{xw}$, $\mathbf{\Phi}_{xe}$, $\mathbf{\Phi}_{uw}$, and $\mathbf{\Phi}_{ue}$ are computable \textit{given} $\myK$.}

We next focus on the opposite direction: we present necessary and sufficient conditions for the existence of a lower triangular block-matrix $\myK$ that always satisfies \cref{eq_closedloop_mapping},  providing how to compute such a $\myK$ given $\mathbf{\Phi}_{xw}$, $\mathbf{\Phi}_{xe}$, $\mathbf{\Phi}_{uw}$, and $\mathbf{\Phi}_{ue}$, instead of the other way around.

\begin{proposition}[System Level Synthesis for Partially-Observed LTV Systems]\label{prop_SLS_LTVwithMeasurements} 
There exists a lower triangular block-matrix $\mathbf{\mathcal{K}}$ such that \cref{eq_closedloop_mapping} holds true if and only if $\mathbf{\Phi}_{xw}$, $\mathbf{\Phi}_{xe}$, $\mathbf{\Phi}_{uw}$, and $\mathbf{\Phi}_{ue}$ are:
\begin{itemize}
    \item lower triangular block-matrices; and
    \item lie in the affine subspace 
        \begin{subequations}
            \vspace{-2mm}
            \begin{align}
                   \left[\begin{array}{cc}
                        \mathbf{I} - \mathcal{Z} \mathcal{A}   & - \mathcal{Z} \mathcal{B} 
                   \end{array} \right] \mathbf{\Phi} &= \left[\begin{array}{cc}
                                                            \mathbf{I}  &  \mathbf{0} 
                                                       \end{array} \right], \label{eq_Phi_affine_1} \\
                    \mathbf{\Phi} \left[\begin{array}{c}
                        \mathbf{I} - \mathcal{Z} \mathcal{A}  \\
                            - \mathcal{C}
                                   \end{array} \right] &= \left[\begin{array}{c}
                                                                \mathbf{I}  \\
                                                                \mathbf{0} 
                                                           \end{array} \right].\label{eq_Phi_affine_2}
            \end{align}
            \label{eq_Phi_affine}
            \vspace{-3mm}
        \end{subequations}
\end{itemize}

\noindent Also, $\myK$ is computed given $\mathbf{\Phi}_{xw}$, $\mathbf{\Phi}_{xe}$, $\mathbf{\Phi}_{uw}$, and $\mathbf{\Phi}_{ue}$ via:
\begin{equation}\label{eq:calK-fromPhi}
    \mathbf{\mathcal{K}} = \mathbf{\Phi}_{ue} - \mathbf{\Phi}_{uw} \mathbf{\Phi}_{xw}^{-1} \mathbf{\Phi}_{xe}.
\end{equation}
\end{proposition}

{The proof follows similar steps as in \cite[Theorem~2.1]{anderson2019system}. It differs from \cite[Theorem~2.1]{anderson2019system} in the way that the system is partially observed, \ie $\mathbf{y} = \mathcal{C} \mathbf{x} + \mathbf{e}$, and the control input now depends on the measurement models $\mathcal{C}$ and measurement noise, \ie $\mathbf{u} = \mathcal{K} \mathbf{y} = \mathcal{K} \mathcal{C} \mathbf{x} + \mathcal{K} \mathbf{e}$.}

By finding a $\myPhi$ satisfying \Cref{prop_SLS_LTVwithMeasurements}'s constraints, we equivalently find a lower triangular block-matrix $\myK$, and thus a control policy per \cref{eq_ut}, satisfying \cref{eq_closedloop_mapping}.

\subsection{Algorithm for {\Cref{prob_safe_obs_reg_opt}} via SDP Reformulation}\label{subsec:main-result}

\setlength{\textfloatsep}{-0.5mm}
\begin{algorithm}[t]
    \caption{\mbox{Safe Control of Partially-Observed Linear} \mbox{Time-Varying Systems with Minimal Worst-Case Regret}.}
	\begin{algorithmic}[1]
		\REQUIRE Time horizon $T$; system matrices $\{\mathcal{A}, \mathcal{B}, \mathcal{C}\}$; cost matrices $\mathcal{Q}$ and  $\mathcal{R}$; noise's domain sets $\mathbb{W}$ and $\mathbb{E}$; upper bound $r$ to the noise' total magnitude.
		\ENSURE Output-feedback control gains $\mathcal{K}$ for \cref{eq_ut}'s control policy.
		\medskip
		\STATE $\left\{\mathbf{\Phi}_{xw}, \mathbf{\Phi}_{xe}, \mathbf{\Phi}_{uw}, \mathbf{\Phi}_{ue}\right\} \gets$ Solve the Semi-Definite Program in \Cref{theorem_safe_obs_reg_opt}; 
		\STATE $\mathcal{K} \gets \mathbf{\Phi}_{ue} - \mathbf{\Phi}_{uw} \mathbf{\Phi}_{xw}^{-1} \mathbf{\Phi}_{xe}$~{per \Cref{prop_SLS_LTVwithMeasurements}}.
	\end{algorithmic}\label{alg:safe_obs_reg_opt}
\end{algorithm}

We provide an algorithm for \Cref{prob_safe_obs_reg_opt} (\Cref{alg:safe_obs_reg_opt}).  To this end, we reformulate \Cref{prob_safe_obs_reg_opt} as an SDP (\Cref{theorem_safe_obs_reg_opt}). 
We obtain the reformulation via the steps:
\begin{itemize}[leftmargin=9pt]
    \item We change the optimization variables in \Cref{prob_safe_obs_reg_opt} from the output-feedback control gains in $\myK$ to the response matrix $\myPhi$. We thus leverage that finding a feasible $\myPhi$ requires searching over a convex set, in particular, the set defined by \Cref{prop_SLS_LTVwithMeasurements}'s necessary and sufficient conditions  which take the form of linear constraints.  Once a $\myPhi$ is found, then $\myK$ is computed via \cref{eq:calK-fromPhi}.
 
    \item We reformulate the safety constraints in \cref{eq_safety_constraints} as linear matrix inequalities.  To this end, we adopt the dualization procedure introduced in the proof of~\cite[Theorem~3]{martin2022safe};
    
    \item We reformulate \Cref{prob_safe_obs_reg_opt}'s objective function, namely, $\operatorname{worst-case-regret}_T(\mathbf{u})$, as an equivalent minimization problem of a scalar subject to linear matrix inequalities.  To this end, we perform the first step of the proof of \cite[Theorem~4]{goel2021regret} and then apply Schur complement.
\end{itemize}

We use the following notation and definitions to formally state \Cref{prob_safe_obs_reg_opt}'s SDP reformulation and \Cref{alg:safe_obs_reg_opt}:
\begin{itemize}[leftmargin=9pt]
    \item $\calD\triangleq \operatorname { blkdiag }(\calQ,\calR)$, \ie $\calD$ is the diagonal block-matrix whose elements $\calQ$ and $\calR$ define the cost in \cref{eq:cost-basic};

    \item $\mathbf{Z}$ are the dual variables introduced to reformulate the safety constraints in \cref{eq_safety_constraints} as linear inequalities;
    
    \item $\lambda$ is the scalar that once minimized subject to appropriate linear matrix inequalities becomes equal to \Cref{prob_safe_obs_reg_opt}'s objective function, \ie to $\operatorname{worst-case-regret}_T(\mathbf{u})$;

    \item $\myPhi^{c}$ is the response corresponding to the optimal clairvoyant controller in \cref{eq_reg} that ignores the safety constraints;  \ie per~\cite{goel2022measurement,martin2022safe},\footnote{By solving \cref{eq_opt_nc}, $\myPhi^{c}$ can be clairvoyant since \cref{eq_opt_nc} does \textit{not} require $\mathbf{\Phi}_{xw}$, $\mathbf{\Phi}_{xe}$, $\mathbf{\Phi}_{uw}$, and $\mathbf{\Phi}_{ue}$ to be lower triangular block-matrices; that is, the necessary and sufficient condition of \Cref{prop_SLS_LTVwithMeasurements} are \textit{not} both met. Thus, $\myPhi^{c}$ can correspond to a control gain block-matrix $\myK$ that fails to be lower triangular and, as a result, the corresponding control policy $u_t$ will depend on future measurements $\{y_k\}_{k\in\{t+1,\ldots,T\}}$.}
    \begin{equation}
        \begin{aligned}
            \mathbf{\Phi}^{c} \in\; &  \underset{\boldsymbol{\Phi}}{\operatorname{argmin}} \qquad \left\|\,\left[\begin{array}{cc}
            \mathcal{Q}^{\frac{1}{2}} & \mathbf{0} \\
            \mathbf{0} & \mathcal{R}^{\frac{1}{2}}
            \end{array}\right] \mathbf{\Phi} \; \right\|_{F}^{2} \\
        &\! \operatorname{\textit{subject~to}} \;\quad \text{\cref{eq_Phi_affine_1}}, \; \mathbf{\Phi}_{xe}=\mathbf{0}, \;  \mathbf{\Phi}_{ue}=\mathbf{0}.
        \end{aligned}
        \label{eq_opt_nc}
    \end{equation}
\end{itemize}

\begin{theorem} [SDP Reformulation of \Cref{prob_safe_obs_reg_opt}] \label{theorem_safe_obs_reg_opt}
\Cref{prob_safe_obs_reg_opt} is equivalent to the Semi-Definite Program
\begin{subequations}
    \begin{align}
    & \underset{\boldsymbol{\Phi},\, \boldsymbol{Z},\, \lambda}{\operatorname{\emph{min}}}   \quad \lambda \quad \emph{\text{subject to:}}\nonumber\\
    & \ \ \boldsymbol{\Phi}_{xw},\boldsymbol{\Phi}_{xe}, \boldsymbol{\Phi}_{uw}, \boldsymbol{\Phi}_{ue} \text{ being lower block triangular}; \label{eq_safe_obs_reg_opt_prop1} \\
    & \ \ \cref{eq_Phi_affine_1}~and~\cref{eq_Phi_affine_2}; \label{eq_safe_obs_reg_opt_prop2}\\
    & \ \  \mathbf{Z}^\top \left[\begin{array}{c}
         \mathbf{h}_{w} \\
         \mathbf{h}_{e}
    \end{array}\right]  \leq \mathbf{h}, \; \mathbf{H} \mathbf{\Phi} = \mathbf{Z}^\top \left[\begin{array}{cc}
         \mathbf{H}_{w} & \mathbf{0}\\
         \mathbf{0} & \mathbf{H}_{e}
    \end{array}\right] , \; \mathbf{Z}_{ij} \geq 0; \label{eq_safe_obs_reg_opt_dafety_dual}\\
    & \ \ \lambda > 0, \  \left[\begin{array}{cc}
        \mathbf{I} & \calD^{\frac{1}{2}}\mathbf{\Phi} \\
        \mathbf{\Phi}^\top\calD^{\frac{1}{2}}
                & \lambda \mathbf{I} + (\mathbf{\Phi}^{c})^\top
                            \calD
                            \mathbf{\Phi}^{c} \label{eq_safe_obs_reg_opt_schur}
\end{array}\right] \succeq 0.
\end{align}
\label{eq_safe_obs_reg_opt}
\end{subequations}
\end{theorem}

\Cref{theorem_safe_obs_reg_opt} prescribes an SDP in place of \Cref{prob_safe_obs_reg_opt}.
Particularly, \cref{eq_safe_obs_reg_opt} relates to \Cref{prob_safe_obs_reg_opt} as follows: \cref{eq_safe_obs_reg_opt_prop1,eq_safe_obs_reg_opt_prop2} result from the change of the optimization variables in  \Cref{prob_safe_obs_reg_opt} from the output-feedback control gains in $\myK$ to the  response matrix $\myPhi$, per the necessary and sufficient conditions in \Cref{prop_SLS_LTVwithMeasurements}; \cref{eq_safe_obs_reg_opt_dafety_dual} results from the reformulation of the safety constraints in \cref{eq_safety_constraints} as linear inequalities, per the dualization procedure in~\cite[Proof of Theorem~3]{martin2022safe}; and the new objective of minimizing the scalar $\lambda$ subject to \cref{eq_safe_obs_reg_opt_schur} result from the reformulation of \Cref{prob_safe_obs_reg_opt}'s objective function, that is, of $\operatorname{worst-case-regret}_T(\mathbf{u})$.

\myParagraph{\Cref{alg:safe_obs_reg_opt}'s Description}
\Cref{alg:safe_obs_reg_opt} solves \Cref{prob_safe_obs_reg_opt} by (i) solving \Cref{prob_safe_obs_reg_opt}'s equivalent SDP reformulation in \cref{eq_safe_obs_reg_opt} to obtain an optimal  response matrix $\myPhi$ (line 1), and then by (ii) computing the corresponding output-feedback control gain block-matrix $\calK$ per \cref{eq:calK-fromPhi}.


\section{Numerical Evaluations in \\Trajectory Tracking Scenarios}\label{sec:experiments}
We evaluate \Cref{alg:safe_obs_reg_opt} in simulated scenarios of safe control of partially-observed LTV systems for trajectory tracking. We first consider synthetic partially-observed LTV systems aiming to stay at zero despite noise disturbances (\Cref{subsec:sim-1}).  Then, we consider a quadrotor aiming to stay at a hovering position; to this end, the quadrotor collects asynchronous GPS and Inertial Measurement Unit (IMU) measurements (\Cref{subsec:sim-2}).  

\myParagraph{Compared Algorithms} 
We compare \Cref{alg:safe_obs_reg_opt} with the safe $\mathcal{H}_{2}$ and $\mathcal{H}_{\infty}$ controllers \cite{martin2022safe}. 
The clairvoyant controller is obtained by solving \cref{eq_opt_nc}.

\myParagraph{Tested Noise Types}
We corrupt the state dynamics and the sensor measurements with diverse noise: (i) stochastic noise, drawn for the Gaussian, Uniform, Gamma, Exponential, Bernoulli, Weibull, or Poisson distribution, and (ii) non-stochastic noise, in particular, worst-case (adversarial) noise. 

\myParagraph{Summary of Results} \Cref{alg:safe_obs_reg_opt} 
demonstrates a \textit{Best of Both Worlds} (BoBW) performance: either it is better than $\mathcal{H}_2$ and $\mathcal{H}_\infty$ across the tested types of noise, or it performs better than $\calH_2$ or $\calH_\infty$ across all tested types of noise.

\vspace{1mm}
We performed all simulations in MATLAB with YALMIP toolbox \cite{Lofberg2004} and MOSEK solver \cite{mosek}.

Our code will be open-sourced via a link here.

\begin{figure*}[t]
    \captionsetup{font=footnotesize}
	\centering
	\includegraphics[width=0.95\textwidth]{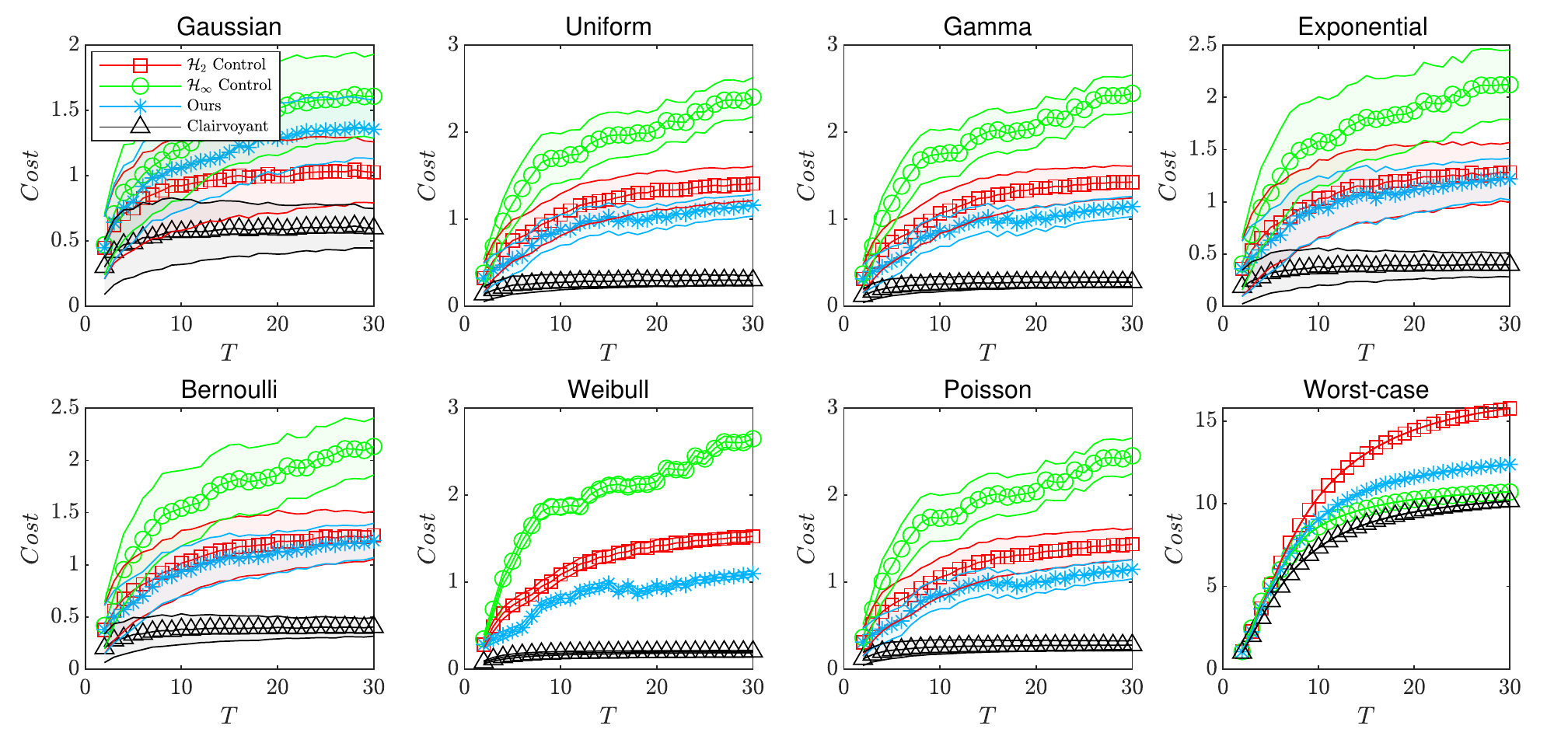}
	\vspace{-3mm}
	\caption{\textbf{Performance Comparison given the LTV System in \Cref{subsec:sim-1} with $\rho = 0.85$ (stable system case).} The performance  is quantified per the cost in \cref{eq:cost-basic}.  Each figure corresponds to a different process and measurement noise type. The shaded areas represent standard deviation.}
	\vspace{-2mm}
    \label{fig_Result_random_085}
\end{figure*} 

\begin{figure*}[t]
	\centering
    \captionsetup{font=footnotesize}
	\includegraphics[width=0.95\textwidth]{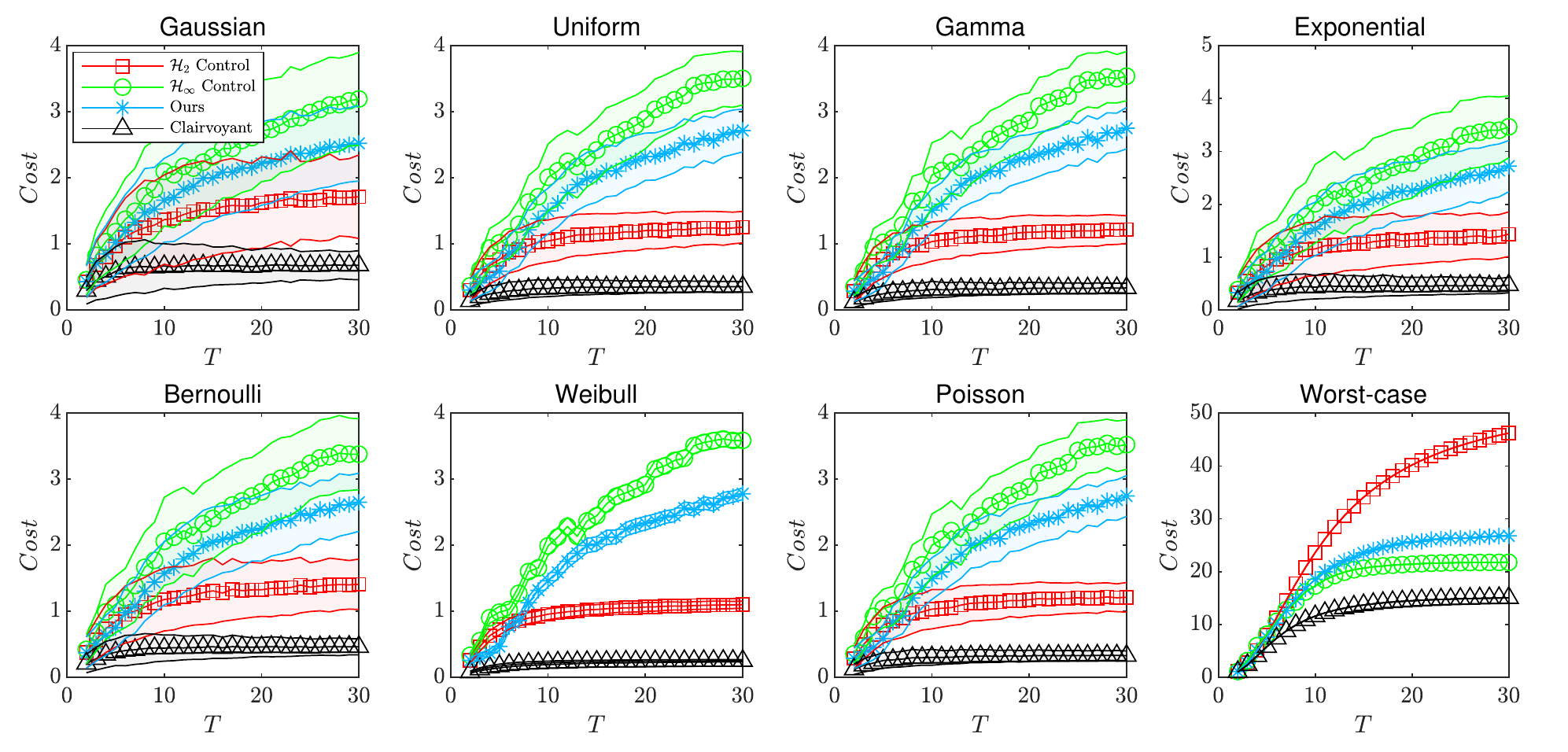}
	\vspace{-3mm}
	\caption{\textbf{Performance Comparison given the LTV System in \Cref{subsec:sim-1} with $\rho = 1.05$ (unstable system case).} The performance  is quantified per  the cost in \cref{eq:cost-basic}.  Each figure corresponds to a different process and measurement noise type. The shaded areas represent standard deviation.}
	\vspace{-2mm}
    \label{fig_Result_random_105}
\end{figure*}

\begin{figure*}[t]
	\centering
	    \captionsetup{font=footnotesize}
	\includegraphics[width=0.95\textwidth]{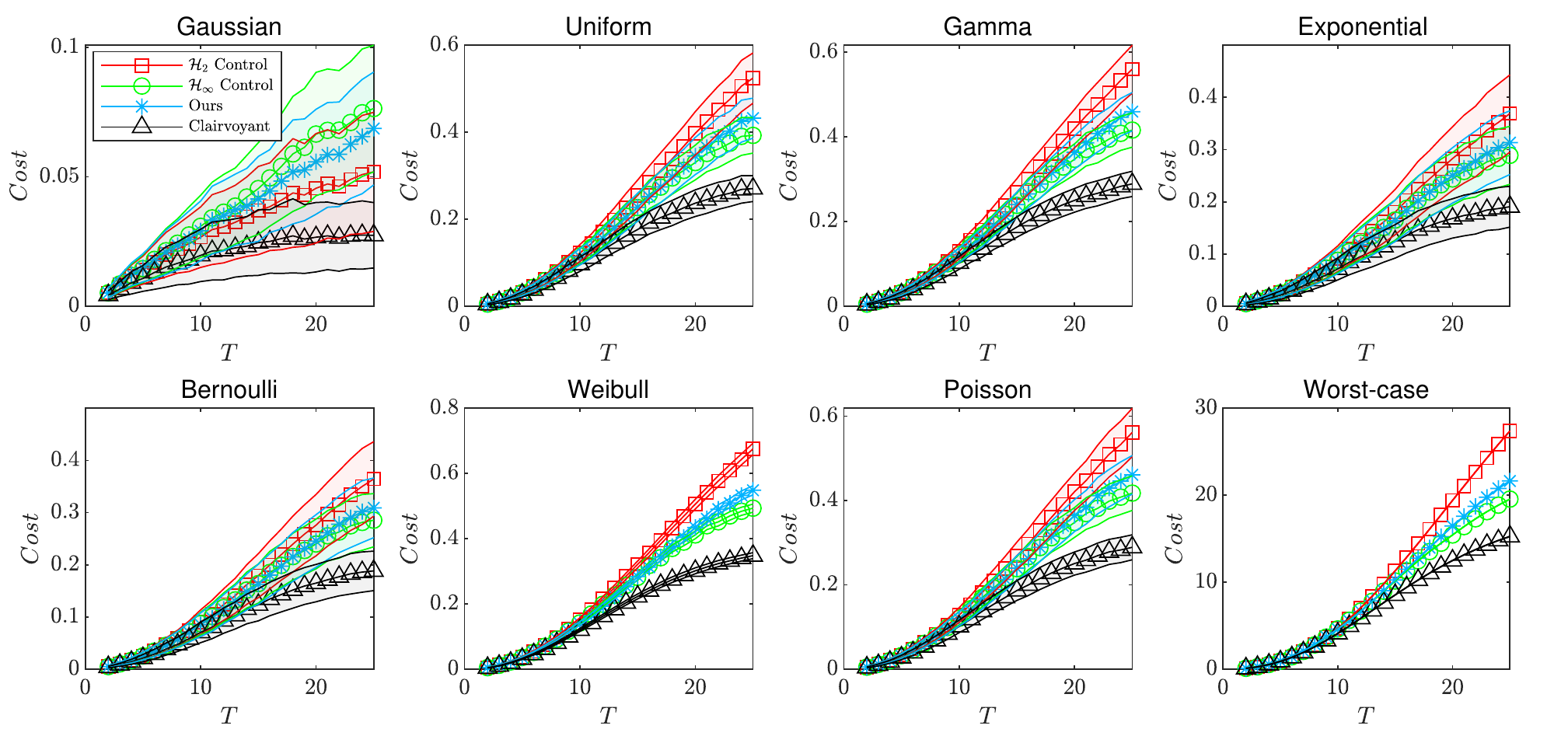}
	\vspace{-3mm}
	\caption{\textbf{Performance Comparison given the Quadrotor System in \Cref{subsec:sim-2}.} The performance  is quantified per the cost in \cref{eq:cost-basic}.  Each figure corresponds to a different process and measurement noise type. The shaded areas represent  standard deviation.}
	\vspace{-5mm}
    \label{fig_Result_UAV}
\end{figure*} 

\subsection{Synthetic Partially-Observed LTV Systems}\label{subsec:sim-1}

\myParagraph{Simulation Setup} We consider LTV systems such that
\begin{equation*}
    \begin{aligned}
            A_{t}&=\rho\left[\begin{array}{ccc}
            0.7 & 0.2 & 0 \\
            0.3 & 0.7 & -0.1 \\
            0 & -0.2 & 0.8
            \end{array}\right], B_{t}=\left[\begin{array}{cc}
            1 & 0.2 \\
            2 & 0.3 \\
            1.5 & 0.5
            \end{array}\right], 
\end{aligned}
\end{equation*}

\begin{equation*}
    \begin{aligned}
            C_{t} &= \left\{ \begin{array}{l}
              \left[\begin{array}{ccc}
            1 & 0 & 0 \\
            0 & 1 & 0 
            \end{array}\right] , \  t = \{1,3,\ldots\};   \\ 
            \  \\
              \left[\begin{array}{ccc}
            0 & 1 & 0 \\
            0 & 0 & 1 
            \end{array}\right] , \ t = \{2,4,\ldots\},             
            \end{array} \right.  
    \end{aligned}
\end{equation*}
where $\rho$ is the spectral radius of the system. 

We demonstrate \Cref{alg:safe_obs_reg_opt} first on an open-loop \textit{stable} system where $\rho=0.85$, and then on an open-loop \textit{unstable} system where  $\rho=1.05$. For the system with $\rho=0.85$, we choose the safety constraints: $-5 \leq x_{t} \leq 5$ and $-5 \leq u_{t} \leq 5$; and we assume noise such that $-1 \leq w_{t} \leq 1$ and $-1 \leq e_{t} \leq 1$.  For the system with $\rho=1.05$, we choose the safety constraints: $-30 \leq x_{t} \leq 30$ and $-30 \leq u_{t} \leq 30$; and we assume noise such that $-1 \leq w_{t} \leq 1$ and $-1 \leq e_{t} \leq 1$. 

We consider that $\mathcal{Q}$ and $\mathcal{R}$ are the identity matrix $\mathbf{I}$. 

We simulate the setting for all $T~\in~\{2, 3, \ldots, 30\}$. 

\myParagraph{Results} The results are summarized in \Cref{fig_Result_random_085} and \Cref{fig_Result_random_105} for $\rho=0.85$ and $\rho=1.05$ respectively. Under Gaussian and worst-case noise, \Cref{alg:safe_obs_reg_opt}'s performance lies between that of the $\calH_2$ and $\calH_\infty$ controllers. Under all other noise types: for $\rho=0.85$, \Cref{alg:safe_obs_reg_opt} outperforms the $\calH_2$ and $\calH_\infty$ controllers; for $\rho=1.05$, \Cref{alg:safe_obs_reg_opt}  outperforms the $\calH_\infty$ controller. In sum, \Cref{alg:safe_obs_reg_opt}  demonstrates a BoBW performance across all scenarios.

\subsection{Hovering Quadrotor}\label{subsec:sim-2}

\myParagraph{Simulation Setup} We consider a quadrotor model with state vector its position and velocity, and control input its roll, pitch, and total thrust. The quadrotors goal is stay at a predefined hovering position.  To this end, we focus on its linearized dynamics, taking the form 
\addtolength{\arraycolsep}{-1.75pt}
\begin{equation*}
    \begin{aligned} 
            A_{t}\!&=\!\left[\begin{array}{cccccc}
            1 & 0 & 0 & 0.1 & 0 & 0 \\
            0 & 1 & 0 & 0 & 0.1 & 0 \\
            0 & 0 & 1 & 0 & 0 & 0.1 \\
            0 & 0 & 0 & 1 & 0 & 0 \\
            0 & 0 & 0 & 0 & 1 & 0 \\
            0 & 0 & 0 & 0 & 0 & 1 
            \end{array}\right], B_{t}\!=\!\left[\begin{array}{ccc}
            -\frac{4.91}{100} & 0 & 0 \\
             0 & \frac{4.91}{100} & 0 \\
             0 & 0 & \frac{1}{200} \\
             -\frac{98.1}{100} & 0	& 0 \\
             0 & \frac{98.1}{100} & 0 \\
             0 & 0 & \frac{1}{10}
            \end{array}\right].
    \end{aligned}
\end{equation*}
\addtolength{\arraycolsep}{1.75pt}

The quadrotor collects GPS and IMU measurements. The GPS measurements are available every $3$ time steps, and the IMU measurements are available in all other time steps, reflecting the real-world scenarios where IMU measurements are more frequently available~\cite{goel2021introduction}. Formally,
\begin{equation*}
    \begin{aligned}
            C_{t} &= \left\{ \begin{array}{l}
              \left[\begin{array}{cccccc}
            1 & 0 & 0 & 0 & 0 & 0 \\
            0 & 1 & 0 & 0 & 0 & 0 \\
            0 & 0 & 1 & 0 & 0 & 0
            \end{array}\right] , \ t = \{1,4,\ldots\},    \\ 
            \  \\
              \left[\begin{array}{cccccc}
            0 & 0 & 0 & 1 & 0 & 0 \\
            0 & 0 & 0 & 0 & 1 & 0 \\
            0 & 0 & 0 & 0 & 0 & 1
            \end{array}\right] , \  t = \{2,3,5,6\ldots\}.              
            \end{array} \right.
    \end{aligned}
    \label{eq_system_UAV}
\end{equation*}

We choose the safety constraints: $-5~\leq~x_{t}~\leq~5$ and $[-\pi \ -\pi \ -20]^\top~\leq u_{t}~\leq~[\pi \ \pi \ 20]^\top$; and we assume noise such that $-0.1 \leq w_{t} \leq 0.1$ and $-0.1 \leq e_{t} \leq 0.1$. 

We consider that  $\mathcal{Q}$ and $\mathcal{R}$ are the identity matrix $\mathbf{I}$. 

We simulate the setting for all $T~=~\{2, 3, \ldots, 25\}$.

\myParagraph{Results} The results are summarized in \Cref{fig_Result_UAV}. \Cref{alg:safe_obs_reg_opt}'s performance lies between that of the  $\calH_2$ and $\calH_\infty$ controllers under Gaussian and worst-case noise. Under all other noise types, \Cref{alg:safe_obs_reg_opt} always outperforms the $\calH_2$ controller, and {is on par with} the $\calH_\infty$ controller up to horizon around $T=15$. All in all, \Cref{alg:safe_obs_reg_opt}  demonstrates a superior or BoBW performance across all scenarios.

\section{Conclusion} \label{sec:con}

\myParagraph{Summary}
We provided an algorithm for the safe control of partial-observed LTV systems against unknown and unpredictable process and measurement noise (\Cref{alg:safe_obs_reg_opt}). \Cref{alg:safe_obs_reg_opt} prescribes an output-feedback control input, guaranteeing safety and minimum worst-case dynamic regret among all noise realizations. To derive \Cref{alg:safe_obs_reg_opt}, we formulated an SDP based on a System Level Synthesis approach for partially-observed LTV systems. We validated \Cref{alg:safe_obs_reg_opt} in simulated scenarios; the algorithm was observed to be better or on par with either the $\mathcal{H}_2$ or $\mathcal{H}_\infty$ controller, demonstrating a \textit{Best of Both Worlds} performance.

\myParagraph{Future Work}
\Cref{alg:safe_obs_reg_opt} plans control policies given a lookahead time horizon, relying on an a priori knowledge of the state, control input, and measurement matrices across the horizon (\Cref{assumption_sys}). This is infeasible in general: \eg in camera-based navigation, where state estimation relies on feature detection and tracking over sequential camera frames, the measurement matrices become known only once the frames have been captured and the features have been detected~\cite{ma2004invitation}. But which will be the frames and which will be the detected features is typically unknown a priori. In our future work, we aim to address the said limitations by enabling online learning variants of our algorithm that (i) do not rely on \Cref{assumption_sys}, and that (ii) provably guarantee a \textit{Best of Both Worlds} performance. We will also ensure its efficient implementation to demonstrate application in real-world systems, in particular, aerial drones that aim to land on moving platforms or perform acrobatics in the presence of unpredictable wind disturbances.

\appendices
\section*{Appendix}\label{Appendix}
\subsection{Proof of \Cref{prop_SLS_LTVwithMeasurements}}\label{proof_prop_SLS_LTVwithMeasurements}

{The proof follows similar steps as in \cite[Theorem~2.1]{anderson2019system}, differing from \cite[Theorem~2.1]{anderson2019system} in that the system is partially observed, \ie $\mathbf{y} = \mathcal{C} \mathbf{x} + \mathbf{e}$, and the control input now depends on the measurement models $\mathcal{C}$ and measurement noise, \ie $\mathbf{u} = \mathcal{K} \mathcal{C} \mathbf{x} + \mathcal{K} \mathbf{e}$.}

We first prove the sufficiency of the conditions in \Cref{prop_SLS_LTVwithMeasurements}, and then their necessity. 

\paragraph*{\textbf{Sufficiency}}
We show that: 1) $\mathbf{\Phi}_{xw}$, $\mathbf{\Phi}_{xe}$, $\mathbf{\Phi}_{uw}$, and $\mathbf{\Phi}_{ue}$ are lower triangular block-matrices; 2) $\mathbf{\Phi}_{xw}$, $\mathbf{\Phi}_{xe}$, $\mathbf{\Phi}_{uw}$, and $\mathbf{\Phi}_{ue}$ lie in the affine space in \cref{eq_Phi_affine}; 3) controller $\calK$ can be recovered from \cref{eq:calK-fromPhi}. Respectively:
\begin{enumerate}[leftmargin=13pt]
    \item \label{app_1_1} The statement holds since $\mathcal{A}, \mathcal{B}, \mathcal{C}$ are block-diagonal, $\mathcal{K}$ is block-lower-triangular, $\mathcal{Z}$ is the block-downshift operator, and the inverse of a block-lower-triangular matrix remains a block-lower-triangular matrix.
    
    \item \label{app_1_2} From \cref{eq_system_compact_y,eq_system_compact_u}, we have
    \begin{equation}
        \mathbf{u} = \mathcal{K} \mathcal{C} \mathbf{x} + \mathcal{K} \mathbf{e}.
        \label{aux_eq_ux}    
    \end{equation}
    \noindent Substituting \cref{aux_eq_ux} into \eqref{eq_system_compact_x} gives:
    \begin{equation}
        \begin{aligned}
              &\mathbf{x} = \mathcal{Z} \mathcal{A} \mathbf{x} + \mathcal{Z} \mathcal{B} \mathcal{K} \mathcal{C} \mathbf{x} + \mathcal{Z} \mathcal{B} \mathcal{K} \mathbf{e} + \mathbf{w} \\
              \Rightarrow \ & \left( \mathbf{I} - \mathcal{Z} \mathcal{A} - \mathcal{Z} \mathcal{B} \mathcal{K} \mathcal{C}\right) \mathbf{x} =  \mathbf{w} + \mathcal{Z} \mathcal{B} \mathcal{K} \mathbf{e} \\
              \Rightarrow \ & \mathbf{x} = \mathbf{\Phi}_{xw} \mathbf{w} + \mathbf{\Phi}_{xe} \mathbf{e}.
        \end{aligned}
        \label{aux_eq_x}
    \end{equation}
Then, substituting \cref{aux_eq_x} into \eqref{aux_eq_ux} gives:
    \begin{equation}
        \begin{aligned}
            \mathbf{u} & = \mathcal{K} \mathcal{C} \mathbf{\Phi}_{xw} \mathbf{w} + \mathcal{K} \mathcal{C} \mathbf{\Phi}_{xe} \mathbf{e} + \mathcal{K} \mathbf{e} \\
            & = \mathcal{K} \mathcal{C} \mathbf{\Phi}_{xw} \mathbf{w} + \left( \mathcal{K} \mathcal{C} \mathbf{\Phi}_{xe} + \mathcal{K}\right) \mathbf{e} \\
            & = \mathbf{\Phi}_{uw} \mathbf{w} + \mathbf{\Phi}_{ue} \mathbf{e}.
        \end{aligned}
        \label{aux_eq_U}
    \end{equation}
From \cref{aux_eq_x} and \cref{aux_eq_U}, we show that \cref{eq_Phi} holds. Now using \cref{eq_Phi}, we show \cref{eq_Phi_affine} holds:
\begin{subequations}
        \begin{align*}
          & \quad \left[\begin{array}{cc}
            \mathbf{I} - \mathcal{Z} \mathcal{A}   & - \mathcal{Z} \mathcal{B} 
          \end{array} \right] \left[\begin{array}{c}
                                \mathbf{\Phi}_{xw} \\
                                \mathbf{\Phi}_{uw}
                              \end{array} \right] \\
          & = \left( \mathbf{I} - \mathcal{Z} \mathcal{A} \right) \mathbf{\Phi}_{xw} - \mathcal{Z} \mathcal{B} \mathcal{K} \mathcal{C} \mathbf{\Phi}_{xw}  \\
          & = \left( \mathbf{I} - \mathcal{Z} \mathcal{A} - \mathcal{Z} \mathcal{B} \mathcal{K} \mathcal{C} \right) \mathbf{\Phi}_{xw} = \mathbf{I}, \\
            & \quad \left[\begin{array}{cc}
                \mathbf{I} - \mathcal{Z} \mathcal{A}   & - \mathcal{Z} \mathcal{B} 
              \end{array} \right] \left[\begin{array}{c}
                                    \mathbf{\Phi}_{xe} \\
                                    \mathbf{\Phi}_{ue}
                                   \end{array} \right]   \\
            & = \left( \mathbf{I} - \mathcal{Z} \mathcal{A} \right) \mathbf{\Phi}_{xw} \mathcal{Z} \mathcal{B} \mathcal{K} - \mathcal{Z} \mathcal{B} \mathcal{K} \mathcal{C} \mathbf{\Phi}_{xw} \mathcal{Z} \mathcal{B} \mathcal{K} - \mathcal{Z} \mathcal{B} \mathcal{K} \\
            & = \left( \mathbf{I} - \mathcal{Z} \mathcal{A} - \mathcal{Z} \mathcal{B} \mathcal{K} \mathcal{C} \right) \mathbf{\Phi}_{xw} \mathcal{Z} \mathcal{B} \mathcal{K} - \mathcal{Z} \mathcal{B} \mathcal{K} \\
            & = \mathcal{Z} \mathcal{B} \mathcal{K} - \mathcal{Z} \mathcal{B} \mathcal{K}  = \mathbf{0}, \\
        & \quad \left[\begin{array}{cc}
            \mathbf{\Phi}_{xw}   & \mathbf{\Phi}_{xe} 
          \end{array} \right] \left[\begin{array}{c}
                                 \mathbf{I} - \mathcal{Z} \mathcal{A}\\
                                - \mathcal{C}
                              \end{array} \right]   \\
        & = \mathbf{\Phi}_{xw} \left( \mathbf{I} - \mathcal{Z} \mathcal{A} \right) - \mathbf{\Phi}_{xw} \mathcal{Z} \mathcal{B} \mathcal{K} \mathcal{C} \\
        & = \mathbf{\Phi}_{xw}  \left( \mathbf{I} - \mathcal{Z} \mathcal{A} - \mathcal{Z} \mathcal{B} \mathcal{K} \mathcal{C} \right)   = \mathbf{I}, \\
        & \quad \left[\begin{array}{cc}
            \mathbf{\Phi}_{uw}   & \mathbf{\Phi}_{ue} 
          \end{array} \right] \left[\begin{array}{c}
                                 \mathbf{I} - \mathcal{Z} \mathcal{A}\\
                                - \mathcal{C}
                              \end{array} \right] \label{aux2_2_1} \\
        & = \mathcal{K} \mathcal{C} \mathbf{\Phi}_{xw} \left( \mathbf{I} - \mathcal{Z} \mathcal{A} \right) - \mathcal{K} \mathcal{C} \mathbf{\Phi}_{xw} \mathcal{Z} \mathcal{B} \mathcal{K} \mathcal{C} - \mathcal{K} \mathcal{C} \\
        & = \mathcal{K} \mathcal{C} \mathbf{\Phi}_{xw} \left( \mathbf{I} - \mathcal{Z} \mathcal{A} - \mathcal{Z} \mathcal{B} \mathcal{K} \mathcal{C}\right) - \mathcal{K} \mathcal{C} \\
        & = \mathcal{K} \mathcal{C} - \mathcal{K} \mathcal{C}= \mathbf{0}.
        \end{align*}
\end{subequations}

    \item  
    By substituting \cref{eq_Phi_uw,eq_Phi_ue}, we get
    \begin{equation*}
        \begin{aligned}
             \mathbf{\Phi}_{ue} - \mathbf{\Phi}_{uw} \mathbf{\Phi}_{xw}^{-1} \mathbf{\Phi}_{xe} 
            & = \mathcal{K} \mathcal{C} \mathbf{\Phi}_{xe} + \mathcal{K} - \mathcal{K} \mathcal{C} \mathbf{\Phi}_{xw} \mathbf{\Phi}_{xw}^{-1} \mathbf{\Phi}_{xe} \\ 
            &= \mathcal{K} \mathcal{C} \mathbf{\Phi}_{xe} + \mathcal{K} - \mathcal{K} \mathcal{C}  \mathbf{\Phi}_{xe} = \mathcal{K} .
        \end{aligned}        
    \end{equation*}
\end{enumerate}

\paragraph*{\textbf{Necessity}}
We show that lower triangular block-matri- ces $\mathbf{\Phi}_{xw}$, $\mathbf{\Phi}_{xe}$, $\mathbf{\Phi}_{uw}$, and $\mathbf{\Phi}_{ue}$  that satisfy \cref{eq_Phi_affine} and \cref{eq:calK-fromPhi} lead to a lower triangular block-matrix $\calK$ per \cref{eq_system_compact}. To this end,  \cref{eq_Phi_affine} can be written as
\begin{subequations}
    \begin{align}
        \left( \mathbf{I} - \mathcal{Z} \mathcal{A} \right) \mathbf{\Phi}_{xw} - \mathcal{Z} \mathcal{B} \mathbf{\Phi}_{uw} &= \mathbf{I}; \label{aux_eq_Phi_affine_11} \\
        \left( \mathbf{I} - \mathcal{Z} \mathcal{A} \right) \mathbf{\Phi}_{xe} - \mathcal{Z} \mathcal{B} \mathbf{\Phi}_{ue} &= \mathbf{0}; \label{aux_eq_Phi_affine_12} \\
        \mathbf{\Phi}_{xw} \left( \mathbf{I} - \mathcal{Z} \mathcal{A} \right)  - \mathbf{\Phi}_{xe} \mathcal{C}  &= \mathbf{I}; \label{aux_eq_Phi_affine_21} \\
        \mathbf{\Phi}_{uw} \left( \mathbf{I} - \mathcal{Z} \mathcal{A} \right)  - \mathbf{\Phi}_{ue} \mathcal{C}  &= \mathbf{0} \label{aux_eq_Phi_affine_22} .
    \end{align}
\end{subequations}
We can obtain $\mathbf{\Phi}_{xw} = \left( \mathbf{I} - \mathcal{Z} \mathcal{A} \right) ^ {-1} \left( \mathbf{I} + \mathcal{Z} \mathcal{B} \mathbf{\Phi}_{uw} \right)$, from \cref{aux_eq_Phi_affine_11}.
Since the matrix $\mathbf{\Phi}_{uw}$ is block-lower-triangular, $\left( \mathbf{I} + \mathcal{Z} \mathcal{B} \mathbf{\Phi}_{uw} \right)$ is invertible. Hence, the inverse of $\mathbf{\Phi}_{xw}$ exists.

Given that $\mathbf{\Phi}_{xw}^{-1}$ exists, we define $\mathcal{K} = \mathbf{\Phi}_{ue} - \mathbf{\Phi}_{uw} \mathbf{\Phi}_{xw}^{-1} \mathbf{\Phi}_{xe}$. This ensures $\mathcal{K}$ is block-lower-triangular.

Now we show that \cref{eq_Phi} holds. Firstly, substituting \cref{aux_eq_Phi_affine_21,aux_eq_Phi_affine_22} into the definition of $\mathcal{K}$ gives
\begin{equation}
    \begin{aligned}
        \mathcal{K}\mathcal{C}  &= \mathbf{\Phi}_{ue}\mathcal{C}  - \mathbf{\Phi}_{uw} \mathbf{\Phi}_{xw}^{-1} \mathbf{\Phi}_{xe}\mathcal{C}  \\
        &= \mathbf{\Phi}_{uw} \left( \mathbf{I} - \mathcal{Z} \mathcal{A} \right) - \mathbf{\Phi}_{uw} \mathbf{\Phi}_{xw}^{-1} \left( \mathbf{\Phi}_{xw} \left( \mathbf{I} - \mathcal{Z} \mathcal{A} \right) - \mathbf{I}  \right)  \\
        &= \mathbf{\Phi}_{uw} \mathbf{\Phi}_{xw}^{-1} \quad \Rightarrow \quad
       \mathbf{\Phi}_{uw} = \mathcal{K}\mathcal{C}\mathbf{\Phi}_{xw}.
    \end{aligned}
    \label{aux_eq_Phi_xw_uw}
\end{equation}

\noindent Combining \cref{aux_eq_Phi_affine_11,aux_eq_Phi_xw_uw}, we obtain
\begin{equation}
    \begin{aligned}
        &\left( \mathbf{I} - \mathcal{Z} \mathcal{A} \right) \mathbf{\Phi}_{xw} - \mathcal{Z} \mathcal{B} \mathcal{K}\mathcal{C} \mathbf{\Phi}_{xw} = \mathbf{I} \\
        \Rightarrow  \quad & \mathbf{\Phi}_{xw} = \left( \mathbf{I} - \mathcal{Z} \mathcal{A} - \mathcal{Z} \mathcal{B} \mathcal{K} \mathcal{C}\right)^{-1} .
    \end{aligned}
    \label{aux_eq_Phi_xw}
\end{equation}

To find $\mathbf{\Phi}_{ue}$ in terms of $\mathbf{\Phi}_{xe}$, substituting \cref{aux_eq_Phi_xw_uw} into the definition of $\mathcal{K}$ gives,
\begin{equation}
    \begin{aligned}
        \mathbf{\Phi}_{ue} &=  \mathbf{\Phi}_{uw} \mathbf{\Phi}_{xw}^{-1} \mathbf{\Phi}_{xe} + \mathcal{K} = \mathcal{K}\mathcal{C}\mathbf{\Phi}_{xe} + \mathcal{K} ,
    \end{aligned}
    \label{aux_eq_Phi_xw} 
\end{equation} 
which also impies $\mathcal{Z} \mathcal{B} \mathbf{\Phi}_{ue} = \mathcal{Z} \mathcal{B} \mathcal{K}\mathcal{C}\mathbf{\Phi}_{xe} + \mathcal{Z} \mathcal{B} \mathcal{K}$.
Combining it with \cref{aux_eq_Phi_affine_12}, we obtain $\mathbf{\Phi}_{xe}$:
\begin{equation}
    \begin{aligned}
        & \left( \mathbf{I} - \mathcal{Z} \mathcal{A} \right) \mathbf{\Phi}_{xe} = \mathcal{Z} \mathcal{B} \mathcal{K}\mathcal{C}\mathbf{\Phi}_{xe} + \mathcal{Z} \mathcal{B} \mathcal{K} \\
        \Rightarrow & \ \left( \mathbf{I} - \mathcal{Z} \mathcal{A} - \mathcal{Z} \mathcal{B} \mathcal{K}\mathcal{C}\right) \mathbf{\Phi}_{xe} = \mathcal{Z} \mathcal{B} \mathcal{K} \
        \Rightarrow  \ \mathbf{\Phi}_{xe} = \mathbf{\Phi}_{xw} \mathcal{Z} \mathcal{B} \mathcal{K}.
    \end{aligned}
    \label{aux_eq_Phi_xw_xe}
\end{equation}  

To show \cref{eq_system_compact}, we substitute \cref{eq_Phi} into \cref{eq_closedloop_mapping}. \qed

\subsection{Proof of \Cref{theorem_safe_obs_reg_opt}}\label{proof_theorem_safe_obs_reg_opt}
\addtolength{\arraycolsep}{-1.75pt}
We follow the sketch of proof presented in \Cref{subsec:main-result}. Since the LTV output-feedback control law $\mathbf{u}$ can be obtained from $\left\{\mathbf{\Phi}_{xw}, \mathbf{\Phi}_{xe}, \mathbf{\Phi}_{uw}, \mathbf{\Phi}_{ue}\right\}$ per \Cref{prop_SLS_LTVwithMeasurements}, the optimization problem \eqref{eq_prob_safe_obs_reg_opt} can be formulated to optimize over the generalized response. Formally,
\begin{subequations}
\vspace{-3mm}
    \begin{align}
    & \min_{\boldsymbol{\Phi},\, \boldsymbol{Z},\, \lambda}   \quad \operatorname{worst-case-regret}_T(\mathbf{u}) \quad {\text{subject to:}}\nonumber\\
    & \ \ \boldsymbol{\Phi}_{xw},\boldsymbol{\Phi}_{xe}, \boldsymbol{\Phi}_{uw}, \boldsymbol{\Phi}_{ue} \text{ \textit{being lower block triangular}}; \\
    & \ \ \textit{\text{\cref{eq_Phi_affine_1}~and~\cref{eq_Phi_affine_2}}}; \\
    & \ \  \mathbf{H} \mathbf{\Phi} \left[\begin{array}{c}
                                                 \mathbf{w} \\
                                                 \mathbf{e}
                                            \end{array}\right]  \leq \mathbf{h}, \ \forall \mathbf{w}, \mathbf{e}: \left[\begin{array}{cc}
                                                 \mathbf{H}_{w} & \mathbf{0}\\
                                                 \mathbf{0} & \mathbf{H}_{e}
                                            \end{array}\right] \left[\begin{array}{c}
                                                                         \mathbf{w} \\
                                                                         \mathbf{e}
                                                                    \end{array}\right]  \leq  \left[\begin{array}{c}
                                                                                                    \mathbf{h}_{w} \\
                                                                                                    \mathbf{h}_{e}
                                                                                              \end{array}\right] \label{eq_prob_safe_obs_reg_opt_Phi_safety}
\end{align}
\label{eq_prob_safe_obs_reg_opt_Phi}
\end{subequations}

\noindent{where by substituting \cref{eq_closedloop_mapping} into \cref{eq_dyn_reg} we can rewrite $\operatorname{worst-case-regret}_T(\mathbf{u})$ as}
\begin{align*}
    \max_{\|\mathbf{w}\|_2^2\,+\,\|\mathbf{e}\|_2^2\,\leq\, r^2} & \left[\begin{array}{c}
        \mathbf{w} \\
        \mathbf{e}
        \end{array}\right]^{\top} \left( \mathbf{\Phi}^\top \calD \mathbf{\Phi}  -  (\mathbf{\Phi}^{c})^\top\calD\mathbf{\Phi}^{c} \right) \left[\begin{array}{c}
        \mathbf{w} \\
        \mathbf{e}
        \end{array}\right].
        \label{eq_worst_case_dyn_reg_quad}
\end{align*} 

In light of the first step of the proof of \cite[Theorem~4]{goel2021regret}, we can equivalently write $\operatorname{worst-case-regret}_T(\mathbf{u})$ as
\begin{equation}
  \min_{\lambda\, >\,0}\; \lambda  \;\text{ subject to } \; \lambda \mathbf{I} - \left( \mathbf{\Phi}^\top \calD \mathbf{\Phi}  -  (\mathbf{\Phi}^{c})^\top\calD\mathbf{\Phi}^{c} \right) \succeq 0.
\label{eq_worst_case_dyn_reg_eigen}
\end{equation}
Using Schur complement~\cite{boyd2004convex}, we rewrite the constraint in \cref{eq_worst_case_dyn_reg_eigen} in the form of \cref{eq_safe_obs_reg_opt_schur}. 

The proof is now completed by following the steps of the proof of \cite[Theorem~3]{martin2022safe} to reformulate the safety constraints as a linear matrix inequalities via dualization. 
\qed



\bibliographystyle{IEEEtran}
\bibliography{References.bib}

\end{document}